\documentclass[twocolumn,showpacs,preprintnumbers,amsmath,amssymb]{revtex4}
\usepackage{bm} 
\newcommand\bq{\begin{equation}}
\newcommand\eq{\end{equation}}
\newcommand{\bqn}{\begin{eqnarray}}
\newcommand{\eqn}{\end{eqnarray}}

\begin{document}
\title[Star formation rate density and the stochastic background of
gravitational waves] {Star formation rate density and the
stochastic background of gravitational waves}
\author{Jos\'e C. N. de Araujo$^1$}\email{jcarlos@das.inpe.br}
\author{Oswaldo D. Miranda$^{1,\,2}$}\email{oswaldo@das.inpe.br}
\affiliation{$^1$Divis\~ao de Astrof\'\i sica - Instituto Nacional
de Pesquisas Espaciais, Avenida dos Astronautas 1758 - S\~ao
Jos\'e dos Campos - 12227-010 SP - Brazil}
\affiliation{$^2$Departamento de F\'\i sica - Instituto
Tecnol\'{o}gico de Aeron\'{a}utica, Pra\c{c}a Marechal Eduardo
Gomes 50 - S\~ao Jos\'e dos Campos - 12228-900 SP - Brazil}
\date{\today}
\begin{abstract}
There is in the literature a number of papers addressing the
stochastic background of gravitational waves (GWs) generated by an
ensemble of astrophysical sources. The main ingredient in such
studies is the so called star formation rate density (SFRD), which
gives the density of stars formed per unit time. Some authors
argue, however, that there is, in the equation that determines the
amplitude of the stochastic background of GWs, an additional
$(1+z)$ term dividing the SFRD, which would account for the
effect  of  cosmic  expansion  onto  the   time   variable. We
argue here that the inclusion of this additional term is wrong. In
order to clarify where the inclusion of the $(1+z)$ term is really
necessary, we briefly discuss the calculation of event rates in
the study of GRBs (gamma ray bursts) from cosmological origin.
\end{abstract}
\pacs{04.30.Db, 98.70.Vc, 98.80.-k}
%
\maketitle
\section{STOCHASTIC BACKGROUND OF GRAVITATIONAL WAVES}
Because of the fact the gravitational waves (GWs) are produced by
a large variety of astrophysical sources and cosmological
phenomena it is quite probable that the Universe is pervaded by a
background of such waves.

We have recently shown \cite{deara00} that the dimensionless
amplitude of the stochastic background of GWs produced by an
ensemble of sources is given by
\begin{equation}
h_{\rm BG}^{2} = {1 \over \nu_{\rm obs}}\int h_{\rm single}^{2}
dR, \label{hBG}
\end{equation}
\noindent where $h_{\rm single}$ is the dimensionless amplitude
produced by an event that generates a signal with observed
frequency $\nu_{\rm obs}$; and $dR$ is the differential rate of
production of GWs.

Eq.(\ref{hBG}) is in fact a shortcut to the calculation of
stochastic background of GWs. An interesting characteristic of
this equation is that it is not necessary to know in detail the
energy flux of the GWs produced at each frequency. If the
characteristic values for the dimensionless amplitude and
frequency are known and the event rate is given it is possible to
calculate the stochastic background of GWs produced by an ensemble
of sources.

In particular, for the case of a background produced by an
ensemble of black holes, the differential rate reads
\begin{equation}
dR_{\rm BH} = \dot\rho_{\ast}(z) {dV\over dz} \phi(m)dmdz,
\label{dR_BH}
\end{equation}
\noindent where $\dot\rho_{\ast}(z)$ is the star formation rate
 density [hereafter SFRD (we refer the reader to the Appendix,
where it is shown how the SFRD is obtained from observations); in
$\rm M_{\odot}\,yr^{-1}\,Mpc^{-3}$], $dV$ is the comoving volume
element and $\phi(m)$ is the initial mass function (IMF). In a few
words, the $\phi(m)\,dm$ represents the number of stars per unit
mass in the interval $[m,m+dm]$. The normalization of the IMF is
obtained through the relation
\begin{equation}
\int_{\rm m_l}^{\rm m_u}m\,\phi(m)\,dm = 1,
\end{equation}
\par\noindent where it is usually considered that ${\rm m_l = 0.1
M_{\odot}}$ and ${\rm m_u = 125 M_{\odot}}$ (see, e.g., Ref.
\cite{deara00}).

Some authors argue that there is an additional $(1+z)$ term in the
equation for the differential rate dividing $\dot\rho_\star(z)$,
which would take into account the effect of cosmic expansion onto
the time variable.

In fact, there are two different points of view in the literature
concerning the inclusion or not of the factor $(1+z)$. The first
group does not include this factor dividing the SFRD (see, e.g.,
Refs.\cite{deara00,ferr99,regi01}). On the other hand, the second
group argues that it is necessary to include the factor $(1+z)$ to
account for the time dilation of the observed rate by cosmic
expansion, converting a source-count equation to an event-rate
equation (see, e.g., Refs.\cite{schn00,cowa01,cowa02a,cowa02b}).
We show here why such an additional term does not exist.

In order to address properly this issue, we should note firstly
that the determination of the amplitude of the GWs is obtained
from the flux received by a detector ($F_{\rm GW}$). In this way,
one divides the luminosity of the sources by $4\pi d_{\rm L}^{2}$,
where $d_{\rm L}$ is the so called luminosity distance. The
luminosity obviously refers to the source frame. Since the
luminosity depends on the differential rate, given, for example,
by Eq.(\ref{dR_BH}), the latter also refers to the source frame,
therefore it is not necessary to redshift it. It is worth
stressing that any necessary redshifting is taken into account in
the definition of $d_{\rm L}$.

Another way to be convinced that the $(1+z)$ does not enter into
the calculation of the differential rate of production of GWs,
$dR_{\rm BH}$, is to derive Eq.(\ref{hBG}) through a procedure
completely different from that used in Ref.\cite{deara00}.

Let us write the specific flux received in GWs at the present
epoch as (see, in particular, Eq. (15) in Ref.\cite{farmer03} and
section 12.1 in Ref.\cite{peac99})
\begin{equation}
F_{\nu}(\nu_{\rm obs})= \int \frac{l_{\nu}}{4\pi d_{\rm L}^{2}}
\frac{d\nu}{d\nu_{\rm obs}}dV,
\end{equation}
\noindent where
\begin{equation}
l_{\nu} = \frac{dL_{\nu}}{dV}
\end{equation}
\noindent is the comoving specific luminosity density (given, e.g,
in ${\rm erg\, s^{-1}\, Hz^{-1}\, Mpc^{-3}}$), which obviously
refers to the source frame.

As discussed in Refs.\cite{farmer03,peac99}, the above equations
are valid to estimate a stochastic background radiation received
on Earth independent of its origin. In the present paper $l_{\nu}$
can be written as follows
\begin{equation}
l_{\nu} = \int \frac{dE_{\rm GW}}{d\nu}\dot\rho_{\ast}(z)
\phi(m)dm,
\end{equation}
\noindent where $dE_{\rm GW}/d\nu$ is the specific energy of the
source. Note that in the above equation $\dot\rho_{\ast}(z)$
refers to the source frame, therefore, there is not the putative
$(1+z)$ factor responsible to the time dilation.

Thus, the flux $F_{\nu}(\nu_{\rm obs})$ received on Earth reads
\begin{equation}
F_{\nu}(\nu_{\rm obs})= \int \frac{1}{4\pi d_{\rm L}^{2}}
\frac{dE_{\rm GW}}{d\nu}\frac{d\nu}{d\nu_{\rm
obs}}\dot\rho_{\ast}(z) \phi(m)dmdV.
\end{equation}
Using Eq.(\ref{dR_BH}) it follows that
\begin{equation}
F_{\nu}(\nu_{\rm obs})= \int \frac{1}{4\pi d_{\rm
L}^{2}}\frac{dE_{\rm GW}}{d\nu} \frac{d\nu}{d\nu_{\rm obs}}
dR_{\rm BH}.
\end{equation}
Note that in the above equation, what multiplies $dR_{\rm BH}$ is
nothing but the specific energy flux per unity frequency (in,
e.g., ${\rm erg\, cm^{-2}\, Hz^{-1}}$), i.e.,
\begin{equation}
f_{\nu}(\nu_{\rm obs}) = \frac{1}{4\pi d_{\rm L}^{2}}\frac{dE_{\rm
GW}}{d\nu} \frac{d\nu}{d\nu_{\rm obs}}
\end{equation}
(see, e.g., Ref.\cite{ferr99b}).

On the other hand, the specific energy flux per unit frequency for
GWs is given by Ref.\cite{carr}
\begin{equation}
f_{\nu}(\nu_{\rm obs}) = \frac{\pi c^{3}}{2G}h_{\rm BH}^{2}.
\end{equation}
Also, the spectral energy density, the flux of GWs, received on
Earth, $F_{\nu}$, in ${\rm erg}\,{\rm cm}^{-2}\,{\rm s}^{-1}\,{\rm
Hz}^{-1}$ can be written from Refs.\cite{douglass,hils} as
\begin{equation}
F_{\nu}(\nu_{\rm obs})= \frac{\pi c^{3}}{2G}h_{\rm BG}^{2}\nu_{\rm
obs}.
\end{equation}
From the above equations one obtains
\begin{equation}
h_{\rm BG}^{2} = {1 \over \nu_{\rm obs}}\int h_{\rm BH}^{2}
dR_{\rm BH}, \nonumber
\end{equation}
\noindent which is nothing but the Eq. (\ref{hBG}).

One could argue that what we derive here is nothing but a rehash
of a well known equation for the calculation of electromagnetic
backgrounds, found in textbooks on cosmology, applied to
gravitational radiation backgrounds. But, since some authors in
the GW community are erroneously introducing extra redshift
factors in their calculations, it is worth presenting clearly how
to calculate GW backgrounds of cosmological origin.

Also, these authors do not base their calculations either on the
formulation of electromagnetic backgrounds presented in textbooks
on cosmology or on an equation such as the one derived here. They
inappropriately base their calculations on event rate equations,
in which the time dilation needs to be considered. As a result,
they consider that in other calculations involving the SFRD the
redshifting should be taken into account.

It is worth stressing that the equation for the energy flux of
GWs, we use in our derivation, is not obtained by adapting the
equation for the Poynting flux by replacing the electric field
with the GW amplitude, with the corresponding factors of G and c.
This equation comes from general relativity.

One could argue when it is necessary to take into account the
redshifting related to the time dilation. We now discuss a case
where we must include the factor $(1+z)$ to take into account the
time dilation of the observed rate.

The case is related to the Gamma Ray Bursts (hereafter GRBs),
which are short and intense pulses of $\gamma-$rays, which last
from a fraction of a second to several hundred seconds. It is
worth mentioning that there are in the literature many papers
dealing with the GRB event rates of cosmological origin, namely,
Refs. \cite{porciani01,bromm02,totani02,firmani04}, among others.

The event rate for the GRBs of cosmological origin is related to
the SFRD. The main reason to include the $(1+z)$ factor in this
case is that we observe GRB events over a fixed time window
$\Delta t_{\rm obs}$, which corresponds to $\Delta t_{\rm
obs}/(1+z)$ in the source frame. Thus, the total number of GRBs
can be written as (see, in particular, Ref.\cite{bromm02})
\begin{equation}
N(>z)=\int_{z}^{\infty}\psi_{\rm GRB}(z'){\Delta t_{\rm obs}\over
(1+z')}{dV\over d\,z'}d\,z', \label{grb}
\end{equation}
\noindent where $dV/dz$ is the comoving volume element and
$\psi_{\rm GRB}(z)$ is the number of GRB events per comoving
volume per unit time, which is proportional to
$\dot\rho_{\ast}(z)$.

Generally speaking, in any event rate of cosmological sources, which
involves a formation rate as a function of redshift or the like, the
time dilation must be applied.

We see that the definition presented in Eq.(\ref{grb}) is
completely different from the definition presented in
Eq.(\ref{hBG}). The misuse of the time dilation of the SFRD in the
calculation of the energy flux is due to the fact that this
involves a time rate, as the event rate of cosmological sources
does. But, one has to bear in mind that the luminosity distance,
that links the luminosity (source frame) and the energy flux
(observer frame), is defined in such a way that any redshifting,
including the time dilation, is implicity in its definition.

In the next section, we reinforce the argumentation against the
inclusion of the extra $(1+z)$ term recalling how luminosity, flux
and the luminosity distance are related to each other.

\section{LUMINOSITY, FLUX AND LUMINOSITY DISTANCE}
Although the material present in this section is standard,
appearing in any textbook of cosmology, it is worth having a look
at the relation between luminosity, flux and luminosity distance
in order to recall that any redshifting, in particular the time
dilation, is already taken into account.

An isotropic emission of photons (or gravitons in the present case)
emitted by a source pass through spheres surrounding the source and,
in the absence of expansion, the detected flux is exactly equal to
the fraction of the area of the sphere surrounding the source and
covered by the detector, that is, $dA/4\pi d_{\rm L}^2$ (where $dA$
is the area of the detector) times the luminosity $L$ of the source.
In this case, $d_{\rm L}$ is simply the distance to the source (at
this point we refer the reader to some textbook on cosmology as,
e.g., \cite{peac99,kolbturner,narlikar}).

When the expansion of the Universe is taken into account, this
consideration is slightly modified. In this case, we can use the
Friedmann - Robertson - Walker (FRW) metric as being centered on
the source; because of homogeneity, the comoving distance between
the source and the observer is the same as we would calculate when
we place the origin at the observer's location.

Setting $t={\rm constant}$ and $r={\rm constant}$ in the FRW line
element, then we obtain
\begin{equation}
ds^2=-r^2a^2(d\theta^2+\sin^2\theta d\phi^2),
\end{equation}
\noindent where this equation represents the line element on the
surface of a Euclidean sphere of radius $a\,r$, and $a=a(t)$ is
the scale factor of the Universe.

The total energy emitted by a source per unit time, at the epoch
$t_1$, over the bandwidth ($\nu,\,\nu+\Delta\,\nu$) is given by
\begin{equation}
dL=L\,J(\nu)\,d\nu,\label{energy01}
\end{equation}
\noindent where $J(\nu)$ is the intensity function.

In the case of a source emitting isotropically, when its light
reaches us it is distributed uniformly across a sphere of
coordinates $r=r_1$ centered on the source. However, we should
note that for the source situated at comoving coordinate $r=r_1$
the photons (or gravitons) emitted at time $t_1$ will be detected
at comoving coordinate $r=0$ at time $t_0$.

Note that because of the redshift, the arriving photons (or
gravitons) with frequencies in the range $(\nu_{\rm obs},\,
\nu_{\rm obs}+\Delta\,\nu_{\rm obs})$ left the source in the
frequency range $[\nu_{\rm obs}(1+z),\, (\nu_{\rm
obs}+\Delta\,\nu_{\rm obs})(1+z)]$.

Thus, the quantity of energy that leaves the source (at $r=r_1$)
between the times $t_1$ and $t_1+\Delta\, t_1$ in the above
frequency range is simply obtained by
\begin{equation}
L\,J(\nu_{\rm obs}\,[1+z])\,\Delta\nu_{\rm
obs}(1+z)\,\Delta\,t_1.\label{energy02}
\end{equation}
In order to find the number of photons (or gravitons) that carry
the above quantity of energy, we should divide Eq.
(\ref{energy02}) by the energy $(1+z)\,h\,\nu_{\rm obs}$, where
$h$ is the Planck constant. That is
\begin{equation}
\delta N=\frac{L\,J(\nu_{\rm obs}\,[1+z])\,\Delta \,
t_1}{h\,\nu_{\rm obs}}\,\Delta \nu_{\rm obs}.
\end{equation}
At the time of reception, these photons (or gravitons) are
distributed across a surface area $4\pi\,r_1^2\,a^2(t_0)$ and they
are received over a time interval ($t_0,\,t_0+\Delta\,t_0$). Thus,
the number of particles received by a detector per unit area held
normal to the line of sight and per unit time is given by
\begin{equation}
\frac{L\,J(\nu_{\rm obs}\,[1+z])}{h\,\nu_{\rm
obs}}\,\frac{\Delta\,
t_1}{\Delta\,t_0}\,\frac{1}{4\pi\,r_1^2\,a^2(t_0)}\,\Delta
\nu_{\rm obs}.\label{energy03}
\end{equation}
When received, each photon (or graviton) has been degraded in
energy by the factor $1/(1+z)$. Thus, the photons (gravitons) now
has the energy $h\,\nu_{\rm obs}$. If we multiply Eq.
(\ref{energy03}) by this factor, we obtain the quantity of photons
(gravitons) received per unit time by us across unit proper area
held perpendicular to the line of sight to the source, and over a
bandwidth $(\nu_{\rm obs},\, \nu_{\rm obs}+\Delta\,\nu_{\rm
obs})$. Denoting this quantity $F_{\nu}(\nu_{\rm obs})\,
\Delta\,\nu_{\rm obs}$, we have
\begin{equation}
F_{\nu}(\nu_{\rm obs})\, \Delta\,\nu_{\rm obs}=L\,J(\nu_{\rm
obs}\,[1+z])\,\frac{\Delta\,
t_1}{\Delta\,t_0}\,\frac{1}{4\pi\,r_1^2\,a^2(t_0)}\,\Delta
\nu_{\rm obs}.
\end{equation}
We should note that in the above equation $\Delta\,
t_1/\Delta\,t_0\,=\,1/(1+z)$. This is just the $(1+z)$ factor
claimed by some authors that should divide the SFRD to take into
account the time dilation of the observed rate by cosmic
expansion. As seen below this factor is already incorporated in
the definition of the luminosity distance. Thus, finally we obtain
\begin{equation}
F_{\nu}(\nu_{\rm obs})=\frac{L\,J(\nu_{\rm
obs}\,[1+z])}{4\pi\,r_1^2\,a^2(t_0)\,(1+z)},\label{fluxdensity}
\end{equation}
\noindent where $F_{\nu}(\nu_{\rm obs})$ is the flux density.

In a few words, the above result can be understood as follows: The
photons (gravitons) from the source pass through a sphere of
proper area $4\pi\,r_1^2$ on which the observer sits, where $r_1$
is the comoving distance. Moreover, redshift affects the flux
density in four ways: photon (graviton) energy and arrival rates
are redshifted, reducing the flux by a factor $(1+z)^2$; opposing
this, the bandwidth $\Delta\,\nu$ is reduced by a factor $(1+z)$,
so the energy flux per unit band goes down by one power of
$(1+z)$; finally, the observed photons (gravitons) at frequency
$\nu_{\rm obs}$ were emitted at frequency $\nu_{\rm obs}\,(1+z)$.
Thus, the flux received on Earth from a cosmological source is the
luminosity of the source divided by the total area and divided by
$(1+z)$ as given in Eq. (\ref{fluxdensity}).

The bolometric flux can be obtained from the integration of Eq.
(\ref{fluxdensity}) over all frequencies. The result is
\begin{equation}
F_{\rm bol}=\frac{L_{\rm
bol}}{4\pi\,r_1^2\,a^2(t_0)\,(1+z)^2},\label{fluxbolometric}
\end{equation}
\noindent where the quantity $d_{\rm L}=r_1\,a(t_0)\,(1+z)$ is the
so called luminosity distance. It is worth stressing that Eq.
(\ref{fluxbolometric}) is a direct consequence of the conservation
of energy.

\section{CONCLUDING REMARKS}
The $(1+z)$ dividing the SFRD, which some authors claim would
account for the time dilation of the observed rate by cosmic
expansion, converting a source-count equation to an event-rate
equation (see, e,g, \cite{schn00,cowa01,cowa02a,cowa02b}), is
nothing but the arrival rates, which is implicit in the luminosity
distance. In particular, Refs.\cite{ferr99,regi01} implicitly
corroborate with the argumentation given here.

It is important to stress that for the cosmological background of
GW models we studied in Ref.\cite{araujo-cqg}, the inclusion of
the $(1+z)$ term would reduce by a factor of $\sim 3-5$ the
signal-to-noise ratio (SNR) predicted for the LIGO observatories.
For some models, this would mean to predict a non detection of
such a stochastic background of GWs.

We hope this brief report contributes to clarify the reader why in
the calculation of the {\it amplitude} of a stochastic background
of GWs, from cosmological sources, it is not necessary to redshift
the SFRD as some authors are arguing for; and also why the GRB
community, or any one who calculates cosmological event rates,
should keep using the time dilation in their calculations.
\acknowledgments J.C.N.A. would like to thank the Brazilian agency
CNPq for partial support (grant 303868/2004-0). O.D.M. would like
to thank the Brazilian agency FAPESP for support (grants
02/07310-0 and 02/01528-4). We would like also to thank the
referees for useful comments and suggestions.
\vskip -7pt
\appendix*
\section{THE STAR FORMATION RATE DENSITY}
\vskip -8pt The SFRD is inferred from observations of the light
emitted by stars at various wavelengths. In particular,
\cite{madau96} investigated the galaxy luminosity density of
rest-frame ultraviolet (UV) radiation up to $z\sim 4$, and they
converted it into the SFRD. The rest-frame UV light is considered
to be a direct tracer of star formation because it is mainly
radiated by short-lived massive stars.

These observable samples are flux-limited, and thus the intrinsic
luminosity of the faintest objects in the sample changes with
redshift. This incompleteness of the samples is corrected by using
a functional (Schechter function) to the luminosity function
obtained from the observations themselves. Then, the conversion
from luminosity density to SFRD generally relies on stellar
population models, an assumed star formation history and a
specific choice for the IMF (generally, it is considered the
Salpeter IMF).

In this way, the SFRD is mainly derived from the observed
luminosity at high redshift, that is, $\dot \rho_{\ast}(z)\propto
L$.

\label{lastpage}

\begin{thebibliography}{99}

\bibitem{deara00} J.C.N. de Araujo, O.D. Miranda and O.D. Aguiar,
\prd {\bf 61}, 124015 (2000).

\bibitem{ferr99} V. Ferrari, S. Matarrese and R. Schneider,
Mon. Not. R. Astron. Soc. {\bf 303}, 247 (1999).

\bibitem{regi01} T. Regimbau and J.A. de Freitas Pacheco,
Astron. Astrophys. {\bf 376}, 381 (2001).

\bibitem{schn00} R. Schneider, A. Ferrara, B. Ciardi, V. Ferrari and S.
Matarrese, Mon. Not. R. Astron. Soc. {\bf 317}, 385  (2000).

\bibitem{cowa01} D. M. Coward, R.R. Burman and D.G. Blair,
Mon. Not. R. Astron. Soc. {\bf 324}, 1015 (2001).

\bibitem{cowa02a} D. M. Coward, M.H.P.M. van Putten and R.R. Burman,
Astrophys. J. {\bf 580}, 1024 (2002).

\bibitem{cowa02b} D. M. Coward, R. R. Burman and D. G. Blair
Mon. Not. R. Astron. Soc. {\bf 329}, 411 (2002).

\bibitem{farmer03} A. J. Farmer and E. S. Phinney, Mon. Not. R.
Astron. Soc. {\bf 346}, 1197 (2003).

\bibitem{peac99} J. A. Peacock, in \textit{Cosmological Physics}
(Cambridge University Press, Cambridge, England, 1999).

\bibitem{ferr99b} V. Ferrari, R. Schneider and S. Matarrese,
Mon. Not. R. Astron. Soc. {\bf 303}, 258 (1999).

\bibitem{carr} B. J. Carr, Astron. Astrophys. {\bf 89}, 6 (1980).

\bibitem{douglass}  D. H. Douglass and V. G. Braginsky, in General
Relativity: An Einstein Centenary Survey, edited by S. W. Hawking
and W. Israel (Cambridge University Press Cambridge, England,
1979), p. 90.

\bibitem{hils} D. Hils, P. L. Bender, and R. F. Webbink, Astrophys. J. 360, 75
(1990).

\bibitem{bromm02} V. Bromm and A. Loeb, Astrophys. J. {\bf 575},
111 (2002).

\bibitem{totani02} T. Totani and A. Panaitescu, Astrophys. J. {\bf
576}, 120 (2002).

\bibitem{firmani04} C. Firmani, V. Avila-Reese, G. Ghisellini and
A. V. Tutukov, Astrophys. J. {\bf 611}, 1033 (2004).

\bibitem{porciani01} C. Porciani and P. Madau, Astrophys.
J. {\bf 548}, 522 (2001).

\bibitem{kolbturner} E. W. Kolb and M. S. Turner, in \textit{The Early
Universe} (Addison Wesley Pub. Comp., USA, 1994).

\bibitem{narlikar} J. V. Narlikar, in \textit{Introduction to
Cosmology} (Cambridge University Press, Cambridge, England, 1993).

\bibitem{araujo-cqg} J. C. N. de Araujo, O. D. Miranda and O. D.
Aguiar, Class. Quantum Grav. {\bf 21}, S545 (2004).

\bibitem{madau96} P. Madau, H. C. Ferguson, M. E. Dickinson, M.
Giavalisco, C. C. Steidel and A. Fruchter, Mon. Not. R. Astron.
Soc. {\bf 283}, 1388 (1996).

\end{thebibliography}
\end{document}